\documentclass[usenatbib]{mn2e}
\input epsf
\usepackage[usenames,dvips]{color}
\usepackage{graphicx,natbib}

\title[The X-ray Properties of the Energetic Pulsar PSR J1838-0655]{The X-ray Properties of the Energetic Pulsar PSR J1838-0655}
\author[L. C.-C. Lin, J. Takata, C. Y. Hwang and J. S. Liang]{Lupin Chun-Che Lin$^1$\thanks{E-mail:
lupin@crab0.astr.nthu.edu.tw},  Jumpei Takata$^2$,  Chorng-Yuan Hwang$^1$ and Jau-Shian Liang$^3$\\
$^1$Graduate Institute of Astronomy, National Central University, Jhongli 32001, Taiwan\\
$^2$Theoretical Institute for Advanced Research in Astrophysics, National Tsing Hua University, Hsinchu 30013, Taiwan\\
$^3$Department of Physics, National Tsing Hua university, Hsinchu 30013, Taiwan}

\begin{document}

\date{June 2009; ??? 2009}
\pagerange{\pageref{firstpage}--\pageref{lastpage}}
\pubyear{2009}
\maketitle

\label{firstpage}

\begin{abstract}
We present and interpret several new X-ray features of the X-ray
pulsar PSR~J1838--0655. The X-ray data are obtained from the
archival data of {\it CHANDRA}, {\it RXTE}, and {\it SUZAKU}. We
combine all these X-ray data and fit the spectra with different
models. We find that the joint spectra are difficult to fit with a
single power law; a broken power-law model with a break at around
6.5 keV can improve the fit significantly. The photon index
changes from $\Gamma$ = 1.0 (below $\mathrm{6.5~keV}$) to $\Gamma$
= 1.5 (above $\mathrm{6.5~keV}$); this indicates a softer spectral
behaviour at hard X-rays. The X-ray flux at 2--20 keV is found to
be $\mathrm{1.6\times10^{-11}~ergs~cm^{-2}~s^{-1}}$. The
conversion efficiency from the spin-down luminosity is $\sim
0.9$\% at 0.8--10 keV, which is much higher than that ($\sim
10^{-3}\%$ -- $10^{-4}\%$) of the pulsars that show similar timing
properties. 
We discuss non-thermal radiation
mechanisms for the observed high X-ray conversion efficiency and
find that emission from the magnetosphere of a greatly inclined
rotator is the most favorable interpretation for the conversion rate
and the pulse profiles at X-ray bands. A line feature close to
$\mathrm{6.65~keV}$ is also detected in the spectra of {\it SUZAKU}/XIS;
it might be the K$_\alpha$ emission of highly ionised Fe
surrounding the pulsar.
\end{abstract}


\section{INTRODUCTION}

PSR~J1838--0655 is an X-ray pulsar that was recently discovered. This X-ray
source was first catalogued by {\it Einstein} Imaging Proportional
Counter (IPC) in the Galactic Plane Survey (1E~1835.3--0658;
\citealp{HG88}). {\it ASCA} found that this source, which was
named as AX~J1838.0--0655 according to the nomenclature of ASCA,
is located at the southern edge of the supernova remnant SNR
G25.5+0.0 \citep{BUKY2003}. \citet{Sug2001} suggested
AX~J1838.0--0655 to be a variable source, however,
\citet{BUKY2003} and \citet{Mal2005} re-analysed the {\it ASCA}
data and obtained a steady flux for this source. AX~J1838.0--0655
was finally identified as an X-ray pulsar because of the discovery of the spin period
at 70.5~ms using {\it RXTE} data \citep{GH2008}.

AX J1838.0--0655 is a bright X-ray source with a hard ($\Gamma$ =
0.8 $\pm$ 0.4) non-thermal spectrum \citep{BUKY2003}. The observed
fluxes of {\it ASCA} (0.7--10~keV; \citealp{BUKY2003}), {\it
Chandra} (2--10 keV; \citealp{GH2008}) and {\it Suzaku} (0.4--10
keV; \citealp{AEDB2009}) are $\sim 0.88,~1.1~\rm{and}~1.32\times
10^{-11}~\mathrm{ergs~cm^{-2}~s^{-1}}$, respectively. The flux in 2--6 keV of
the ASCA spectrum ($\sim 0.4\times
10^{-11}~\mathrm{ergs~cm^{-2}~s^{-1}}$) was different from
that of the {\it EXOSAT} source GPS~1835--070 ($\sim 6.3\times
10^{-11}~\mathrm{ergs~cm^{-2}~s^{-1}}$), which was an X-ray source
discovered at the positional uncertainty of AX J1838.0--0655 with
the Medium-Energy (ME) proportional counters on {\it EXOSAT} \citep{War88}. 
However, the variability of the point-source flux between the two
investigations of {\it EXOSAT} and {\it ASCA} might be caused by the
contamination from nearby sources in the detection of {\it EXOSAT}.
The IBIS/ISGRI on board {\it INTEGRAL} has also
detected this X-ray source at hard X-rays/$\gamma$-rays.
\citep{Bass2004,Bird2004}.

A single absorbed power law usually gave a good fit to the X-ray
spectra of PSR~J1838--0655, but the photon indices might vary at
different wavebands. A single power law with a photon index of
$\Gamma = 1.5 \pm 0.2$ and a column density of $N_H = (6.7\pm
1.3)\times 10^{22}~\mathrm{cm^{-2}}$ provided a good fit to the
composite spectra from 1 to 300 keV generated by {\it
ASCA/INTEGRAL} \citep{Mal2005}. At low-energy wavebands, the
observations of {\it ASCA}/GIS and {\it Chandra}/ACIS indicated a
harder spectrum with $\Gamma= 0.8\pm 0.4$ (0.7--10 keV) and
$0.5\pm 0.2$ (2--10 keV) \citep{Sug2001,GH2008}. However, we note
that the photon indices derived from {\it Suzaku}/XIS
\citep[$\Gamma = 1.27\pm0.11$]{AEDB2009} and {\it Swift}/XRT
\citep[$\Gamma = 1.86^{+0.69}_{-0.46}$]{Landi2006} were significantly softer than the
results of {\it ASCA}/GIS and {\it Chandra}/ACIS at similar
wavebands. On the other hand, the results of {\it Beppo-SAX}/PDS,
{\it INTEGRAL}/ISGRI and {\it Suzaku}/HXD showed much softer
spectra at high-energy wavebands with $\Gamma$ around 2.5 to 1.6
\citep{Mal2004,Mal2005,AEDB2009}. Furthermore, \citet{GH2008} and
\citet{AEDB2009} claimed that the spectrum of PSR~J1838--0655
became steepened at around 8--15 keV; the truth and the origin of
the spectral break or steepening are still unclear. Besides, we
note that the detected X-ray emission might contain emission from
the pulsar wind nebula (PWN) ($\Gamma=1.1-2.0$ in 2--10 keV,
\citealp{GH2008}) and/or emission from the supernova remnant.

The high-energy nature of AX J1838.0--0655 has caught more
attention when \citet{Ahar2005} showed a possible connection
between AX J1838.0--0655 and a high-energy TeV source, HESS
J1837--069. \citet{Ahar2006} indicated that HESS J1837--069 might
be associated with supernova remnants or pulsar wind nebulae; and
this suggested that AX J1838.0--0655 might be a pulsar candidate
before its pulsation was confirmed. Furthermore, the non-thermal
spectral behaviour with a photon index similar to the general
pulsars ($\Gamma \sim 0.5-2.5$) and the large flux ratio between
the X-ray and optical band, $F_{0.4-10 \mathrm{keV}}$/$F_{1 \mathrm{eV}}$
(\citealp{GH2008}; this value is $\sim$100--1000 for the Vela pulsar) also
gave supporting evidence indicating this X-ray source to be a pulsar.

No radio counterpart of the X-ray pulsar PSR~J1838--0655 has been
detected yet. If HESS~J1838--069 is the TeV counterpart of the
X-ray pulsar, the radio brightness of HESS~J1838--069 will be
extremely faint comparing with other known TeV sources. Besides,
unlike the Geminga pulsar, PSR~J1838--0655 has no detectable
$\gamma$-rays from hundreds MeV to hundreds GeV \citep{Har99,
Abdo2009}.

The energy conversion efficiency of PSR~J1838--0655 is very unique
comparing with other known X-ray pulsars with similar properties.
For example, the characteristic age, the surface dipole magnetic
field, and the spin-down luminosity of PSR~J1838--0655 are all
very similar to those of PSRs~B0833--45 (the Vela) and B1706--44;
however, the energy conversion efficiency of PSR~J1838--0655 is
about two to three order of magnitude higher than those of
PSRs~B0833--45 and B1706--44 as inferred from the observed X-ray
emission in 0.8--10~keV. Therefore, PSR~J1838--0655 provides a
unique opportunity to investigate the energy conversion mechanism
and to understand the nature of the X-ray emission of pulsars.

To understand the puzzling properties of the X-ray emission of
PSR~J1838--0655, we re-examine the X-ray spectral behaviour of this
pulsar using {\it Chandra}, {\it RXTE} and {\it Suzaku} archives.
We propose three possible mechanisms to explain the observed
X-ray spectrum and the high energy conversion efficiency of
PSR~J1838--0655. We also marginally detect a line feature at
around 6.65~keV, which might be K$\alpha$ emission of highly
ionised Fe on or around the surface of the pulsar.

\section{Observations and Data Analysis}
To study the emission mechanisms of PSR~J1838--0658, we
re-analysed its spectral and temporal behaviours using the
X-ray archives of {\it Chandra}, {\it RXTE} and {\it Suzaku}.

\subsection{{\it Chandra Data}}

The Chandra data were observed on 2006 August 19 using the
Advanced CCD Imaging Spectrometer operating in the Timed Exposure
(TE)/VFAINT mode with an exposure time of 20 ks.  The same data
have been analysed by \citet{GH2008}. The data reduction and
spectral analysis were performed using standard procedures with
the X-ray packages of CIAO (ver. 3.4.1.1), CALDB 4.1.1 and XSPEC
(ver. 12.4.0).  We followed \citet{GH2008} in selecting the
regions for the pulsar, the PWN, and the background. The spectra
of PSR~J1838--0655 and the PWN are grouped to have at least 50
counts for each spectral bin.

\subsection{{\it RXTE Data}}

The {\it RXTE} data were observed from 2008 February 17 to
March 5 using the Proportional Counter Array (PCA; 2--60 keV). The
same data set has been used by \citet{GH2008} to determine the
period and period derivative of PSR~J1838--0655. The time
resolution of the data set is very high (1 $\mu$s) and can provide
a proper detection of the pulsed spectra of the pulsars. We
performed the standard data reduction procedures with the FTOOLS
(ver 6.5) of HEASoft (ver. 6.5.1) using the latest PCA calibration
files (20090313). The events that we used for spectral analysis
were restricted to Good Xenon and Good Time Interval (GTI). The
total exposures of GTIs are 12.9 ks for PCU 0, 9.3 ks for PCU 1,
31.5 ks for PCU 2 , 4.1 ks for PCU 3 and 8.2 ks for PCU 4. We also
set the energy boundary of photons at 2--20 keV.

\begin{figure}
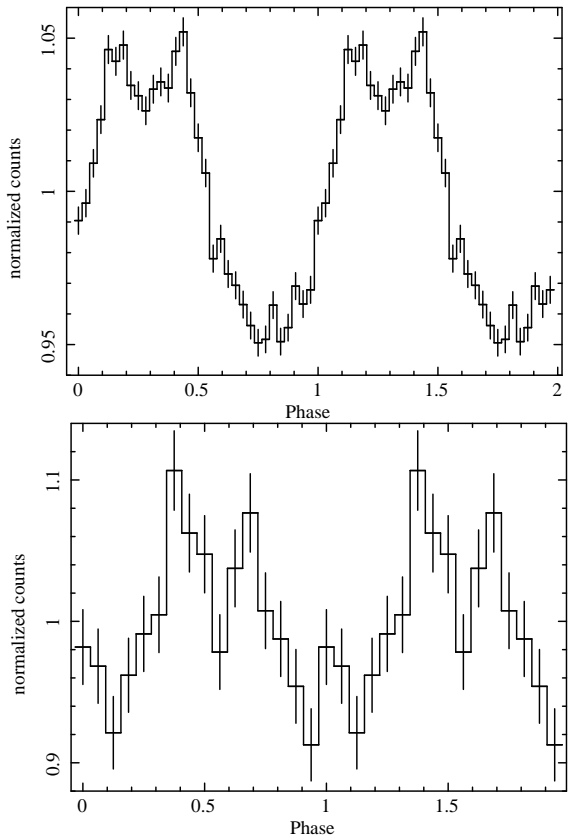

\centering
\includegraphics[width=5.5cm,angle=-90]{2-20_PF.ps}
\includegraphics[width=5.5cm,angle=-90]{10-60_PF.ps}
\caption[Pulse Profile deduced by PCA/RXTE and HXD/Suzaku]{\small (a).The upper panel shows pulse profile in 2--20~keV folded by 32 bins with the period of 70.49824397~ms and the period derivative of $4.925\times 10^{-14}$ using 8 {\it RXTE}/PCA data from 2008 Feb. 17 to Mar. 5. The time zero (t$_0$) of the profile was set at the epoch of 54522.000696574~MJD. (b).The lower panel shows the pulse profile in 10--60 keV folded by 16 bins with the period of 70.496725~ms using {\it Suzaku}/HXD-PIN data of 2007 Mar. 5. The time zero of the profile was set at the start of GTI (54164.53394270~MJD). Both (a). and (b). are divided into two groups to represent the pulsed phase and unpulsed phase. We determined the pulsed phase (on-peak emission) as 0.625 cycle. The pulsed phase of (a). starts from 1st to 20th bin of 32 bins and it of (b). starts from 4th to 14th bin of 16 bins. The plot of the counts are normalized to the average photons in each bin.}
\label{Profile}
\end{figure}

After getting the final products, we performed the solar system barycentric
time correction using ``faxbary'' at (J2000) R.A.=$279^{\circ}.513$,
decl.=$-6^{\circ}.926$ to produce an event time list for following
analysis. According to the timing ephemeris derived by
\citet{GH2008}, we set the period as 70.49824397~ms and the
period derivative $4.925\times 10^{-14}~\mathrm{s\cdot s^{-1}}$ at
the epoch of 54522.000696574~MJD and folded the {\it RXTE} data to
produce the pulsed profile of PSR~J1838--0655 as shown in Fig.~\ref{Profile}.

To generate the pulsed spectrum, we assumed the unpulsed emission
to be the background. We divided the data into ``on-peak'' and
``off-peak'' emission depending on their pulsed phase. The
response of each PCU was produced by the Perl script of ``pcarsp''
and then combined by the task of ``addrmf'' with the weighting
depending on the relative exposures. The pulsed spectrum was then
obtained by subtracting the off-peak emission from the on-peak
emission.

\subsection{{\it Suzaku Data}}

The {\it Suzaku} observations of PSR~J1838--0655 were carried out in
March of 2007. The archival data contain observational results of
the X-ray Imaging Spectrometer (XIS) and the non-imaging Hard
X-ray Detector (HXD). The data sets have been analysed by
\citet{AEDB2009}.

\subsubsection{{\it XIS}}

The {\it Suzaku}/XIS observations of PSR~J1838--0655 have a total
exposure of 42.2 ks. The data were observed in the normal mode
without window option and the pixels on the CCD were read out
every 8~sec. The 42.2~ks exposure of XIS observation was divided
into two editing modes of 5$\times$5 and 3$\times$3 with 14.9~ks
and 27.3~ks respectively.

The selection of source area for the {\it Suzaku}/XIS data in our
analysis is slightly different from that of \citet{AEDB2009}.
Because PSR~J1838--0655 is contaminated by some nearby sources,
which can marginally be resolved in the XIS image (e.g. GPSR5
25.252--0.139; \citealp{GH2008}), we extracted the spectra of
PSR~J1838--0655 only within $1'.5$ circular regions centered at
(J2000) R.A.=$18^h38^m03^s.13$, decl.=$-06^{\circ}55'33''.4$
instead of the regular adoption of $3'$ circular region for a
point source. Our selection still contains $\sim$ 70\% of the
total energy for a point source. We extracted the background of
the pulsar from a concentric annulus of radii $2'.5 < r \leq 3'.5$
centered at the pulsar position and set the energy boundary to be
0.5--10 keV. The extracted spectra comprise 2935 counts for XIS 0,
2180 counts for XIS 1 and 3427 counts for XIS 3 after subtracting
the background.

The XIS data reduction and spectral analysis were executed using
XSELECT (ver. 2.4a) of HEASoft (ver. 6.5.1) and XSPEC (ver.
12.4.0) with the latest {\it Suzaku}/XIS calibration files
(20090203). We generated the response matrix (rmf) and auxiliary
response (arf) files with the HEASoft command of ``xisrmfgen'' and
``xissimarfgen''. The spectra are produced with each channel
containing at least 50 counts. A cross-calibration term was included to
correct the difference among three XIS spectra at 0.5--10 keV.


\subsubsection{{\it HXD}}

The {\it Suzaku}/HXD observations of PSR~J1838--0655 have a total
exposure of 37.7 ks and offer a time resolution of 6~$\mu$s for pulsed detection.
However, the spectrum of the {\it Suzaku}/GSO (Gadolinium Silicate phoswich counters) 
data are not good enough for statistical analysis, we
thus only reduced and analysed the pulsed spectral data obtained
from the HXD-PIN (Positive Intrinsic Negative silicon diodes) data. We performed the data reduction and
spectral analysis procedures of the PIN data using XSELECT
(ver. 2.4a), FTOOLS (ver 6.5) of HEASoft (ver. 6.5.1) and XSPEC
(ver. 12.4.0) with the latest HXD calibration Files (20090203).
The events were restricted to the effective energy range (10--60
keV) of the PIN detector. We also applied solar system
barycentric time correction with the task ``aebarycen'' at (J2000)
R.A.=$279^{\circ}.513$, decl.=$-6^{\circ}.926$ in order to produce
and analyse the pulsation of PSR~J1838--0655. Based on the timing
ephemeris derived by \citet{GH2008}, we set a trial period of
70.496725~ms at epoch MJD 54164.98049 (the mid-point of this
{\it Suzaku}/HXD observation).

The hard X-ray spectrum obtained by subtracting the non-X-ray
background and cosmic X-ray background was divided into two groups
of phase cycles to represent the ``on-peak'' and ``off-peak''
emission. The pulsed spectral data were obtained by subtracting
the off-peak emission from the on-peak emission and rebinned to
ensure that the photons in each spectral channel are larger than
50 after subtracting the un-pulsed background. The pulsed photons
of PSR~J1838--0655 above 50 keV are very few in the data and are
thus ignored in spectral analysis.

\begin{table*}
\begin{minipage}{165mm}
\caption{Single Power-law spectra for AX~J1838.0--0655/PSR~J1838-0655 detected by different X-ray observations.}
\label{NTSpectra}
\begin{tabular}{@{}cccccc}
\hline
\hline
Instrument & Energy range & Photon index & Column density & Unabsorbed flux & References \\
  & (keV) & ($\Gamma$) & (10$^{22}$ cm$^{-2}$) & (10$^{-11}$ ergs cm$^{-2}$ s$^{-1}$) & \\
\hline
{\it Einstein}/IPC & 0.2--3.5 & --- & --- & 0.014$^a$ & \citet{HG88} \\
{\it ASCA}/GIS &  0.7--10.0 & 0.8 (0.4--1.2) & 4.0 (2.8--5.7) & 1.1 & \citet{Sug2001} \\
{\it Beppo-SAX}/PDS & 20--100 & 2.5 (2.2--2.8) & --- & 3.6 & \citet{Mal2004}\\
{\it INTEGRAL}/ISGRI & 20--300 & 1.66 (1.43--1.89) & --- & 9.0 & \citet{Mal2005} \\
{\it Swift}/XRT & 0.2--8.5 & 1.86 (1.4--2.55) & 5.54 (3.73--8.37) & 1.13$^b$ & \citet{Landi2006} \\
{\it Chandra}/ACIS & 2--10 & 0.5 (0.3--0.7) & 4.5 (3.7--5.2) & 0.88 & \citet{GH2008} \\
  & & 0.5 (0.2--0.7) & 4.1 (2.7--5.5) & 0.86 & This work \\
{\it RXTE}/PCA$^c$ & 2--20 & 1.2 (1.1--1.3) & 4.5 (fixed) & 0.9$^b$ & \citet{GH2008} \\
  & & 1.3 (1.2--1.4) & 4.1 (fixed) & 0.92$^b$ & This work \\
{\it Suzaku}/XIS & 0.4--10 & 1.27 (1.16--1.38) & 5.4 (4.9--5.9) & 1.32 & \citet{AEDB2009} \\
  & 0.5--10 & 1.37 (1.28--1.50) & 6.0 (5.6--6.5) & 1.26 & This work \\
{\it Suzaku}/HXD$^c$ &12--50 & 2.0 (1.1--3.0) & --- & 1.84 & \citet{AEDB2009} \\
  & 10--50 & 2.0 (0.9--3.3) & 6.7 (fixed) & 1.9 & This work \\
\hline
\multicolumn{6}{l}{Notes: The errors are set in 90\% confidence interval throughout}\\
\multicolumn{6}{l}{$a$) X-ray flux in IPC counts s$^{-1}$. $b$) Indicated in 2--10 keV. $c$) Only the pulsed spectra were calculated.}
\end{tabular}
\medskip
\end{minipage}
\end{table*}

\section{Results}

Table~\ref{NTSpectra} shows the results of our spectral analysis
for {\it Chandra}, {\it RXTE}, and {\it Suzaku} observations. We
also show the results of previous analysis in Table 1 for
comparisons. Our results are consistent with previous
analyses; however, some significant discoveries are found in
our work. First, we identify a spectral break at $6.5\pm1.0$ keV, which
is roughly consistent with the previous claim of \citet{GH2008} and \citet{AEDB2009} that the spectrum becomes steepen at around 8-15 keV. 
Second, we find a line feature
around 6.4 keV in the spectra of {\it Suzaku}/XIS.

We have applied a simultaneous fit to a composite spectrum of
PSR~J1838--0655. We included the spectra of ACIS/{\it Chandra},
PCA/{\it RXTE} and HXD/{\it Suzaku}; all these spectra are
believed to be dominated by the pulsed phases. To account for the
cross-calibration mismatch between each instrument, we also
introduced a constant to this fit. In order to compare with the
the joint spectrum of {\it ASCA/INTEGRAL} data \citep{Mal2005}, we
have fixed the column density to $6.7\times
10^{22}~\mathrm{cm^{-2}}$ in our analysis. A single absorbed
power-law model for the composite spectrum have a power-law index
of $\Gamma= 1.2\pm 0.1$ with $\chi^2_{\nu}=
1.35~\rm{for}~91~\rm{dof}$. The statistics does not improve even
we set the absorbed column density as the free parameter to fit
the spectra.

A broken power-law model can obviously improve the fit of the joint spectrum
and the additional power-law component is significant at more than 99\% via a F-test. 
The power-law indices change from 1.0$^{+0.1}_{-0.2}$ to 1.5$^{+0.3}_{-0.2}$
with the broken energy at 6.5$^{+1.0}_{-1.0}$ KeV. The statistics
have significantly improved to $\chi^2_{\mu}=
1.20~\rm{for}~89~\rm{dof}$. We note that a similar spectral break
have been discovered in the Crab pulsar \citep{Kup2001}.

\begin{figure}
\centering
\includegraphics[width=6.0cm,angle=-90]{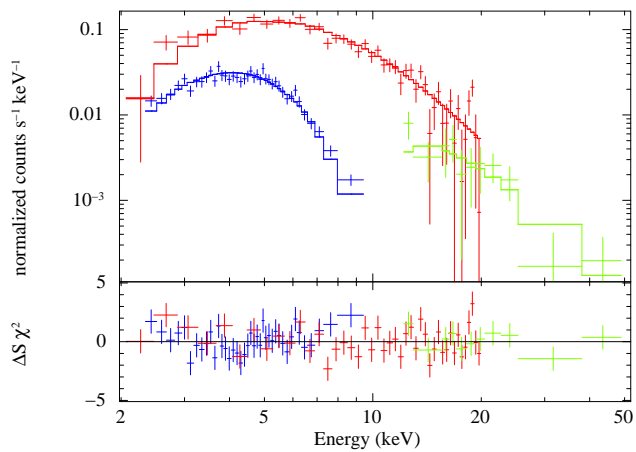}
\caption[Investigation of the Spectral Break]
{\small Cross-calibration fit of the joint spectra (Blue: {\it Chandra}/ACIS; Red: {\it RXTE}/PCA and Green: {\it Suzaku}/HXD-PIN; The vertical axis of the {\it RXTE} spectrum is referred to only one PCU). We applied a broken power-law with a change of the slope from 1.0 to 1.5 and a break energy at 6.5 keV to get the best spectral fit. The bottom panel shows the residuals in terms of $\sigma$s with error bars of size one.}
\label{Spbreak}
\end{figure}

To check whether the broken power-law is caused by the
instrumental responses of different observations, we have also
applied the power-law models to different energy bands of the {\it
RXTE} data to examine the variation of the photon indices of the
pulsed spectrum. Originally, no significant variation of the
photon index was found between the energy domains less than 10~keV
and larger than 10~keV ($\Gamma \sim 1.0-1.1$). However, we note
that both photon indices are flatter than that of the total
spectrum. We thus examine the photon index of the spectrum at the
energy of 5-15~keV and find the photon index $\Gamma \sim 1.4$.
This suggests that there is a spectral break at around 10~keV, and
the break is not caused by the uncertainty of the response
functions among the different instruments.

We also note that the {\it RXTE} spectrum is very different from
the {\it Chandra} one, which shows a much flatter photon index
($\Gamma \sim 0.5$). We note that the time resolution for the {\it
Chandra} TE mode data is only about 3.2 sec, and it is not
adequate to get the pulsed spectrum from timing analysis. However,
if the un-pulsed emission is mainly contributed by the surrounding
ionised cloud, the pulsed spectrum can roughly be obtained by
considering the spectrum of its PWN as the background of the
pulsar. The PWN spectrum can be well fitted with a power law with
a photon index of $\Gamma \sim 1.5 (1.1-2.0)$, and the flux of the
PWN is $\sim 1.6\times 10^{-12} \mathrm{ergs~cm^{-2}~s^{-1}}$. The
flux ratio between PSR~J1838--0655 and its PWN is $> 5.3$ in our
calculation, and this ratio is at the high end of the distribution
of PWNe observed by {\it Chandra} \citep{KP2008}. This suggests
that the observed pointed-source spectrum of {\it Chandra} is
mainly from the pulsar even we do not have an appropriate time
selection. This suggests that the un-appropriate time selection
can not be the cause of the different the photon indexes between
{\it RXTE} and {\it Chandra} spectra. The origin of the
difference is still unclear; but since the {\it Chandra} and {\it
RXTE} observations were carried out at different epochs, the
difference might indicate that the photon index could vary with
time if the difference is not caused by the uncertainty of the
response functions between the different instruments.

\begin{figure}
\centering
\includegraphics[width=6.0cm,angle=-90]{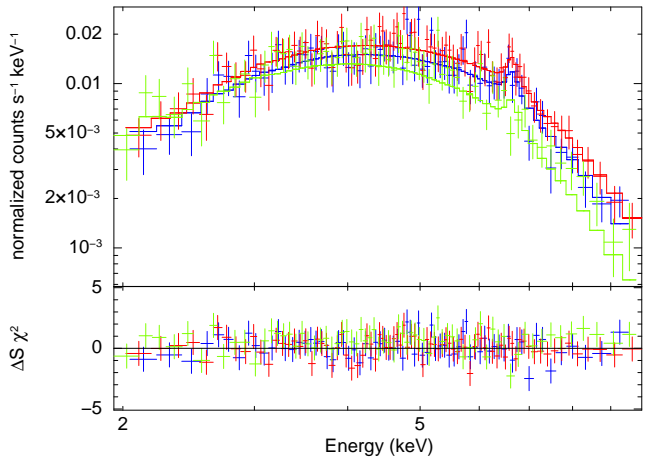}
\caption[Line feature detected by Suzaku/XIS]{\small Spectra of 3 XIS detectors from 2-10 keV (Blue: XIS 0; Green: XIS 1 and Red: XIS 3). Each spectral channel has at least 30 counts. In the simultaneous fitting for the spectra of two FI XIS detectors, we proposed the absorbed power-law and Lorentz line profile to fit the data sets and the line emission was marginally detected close to 6.4 keV. The bottom panel shows the residuals in terms of $\sigma$s with error bars of size one.}
\label{line}
\end{figure}
\begin{figure}
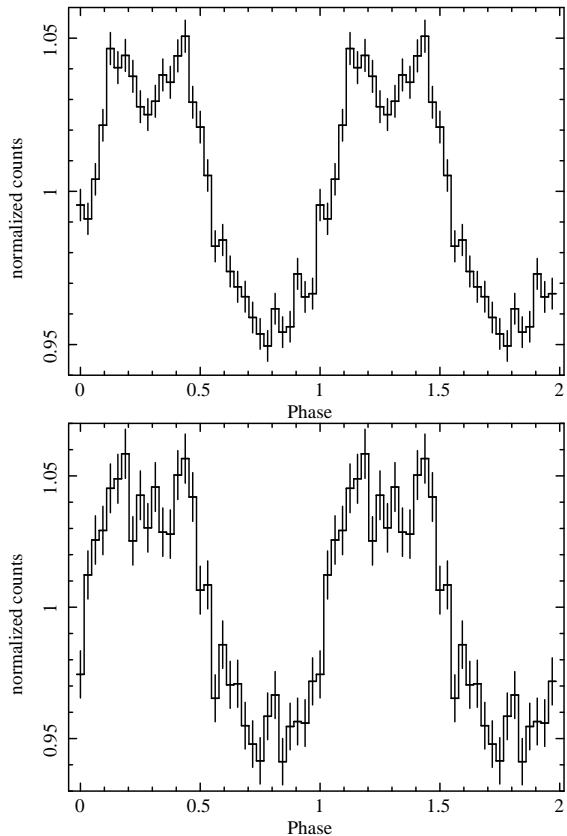

\centering
\includegraphics[width=5.5cm,angle=-90]{2-10_PF.ps}
\includegraphics[width=5.5cm,angle=-90]{10-20_PF.ps}
\caption[Pulse Profile deduced by PCA/{\it RXTE} at different energy bands]{\small Pulse profile of {\it RXTE}/PCA observations folded at different energy bands. All the input parameters are corresponding to Fig.~\ref{Profile}. The upper panel shows the folded light curve of PSR~J1838--0655 at 2--10~keV with $\sim 70.5$~ms and the lower panel shows similar folded profile for the same pulsar at 10--20~keV with the same period. The plot of the counts are also normalized to the average photons in each bin.}
\label{RXTEProfile}
\end{figure}

A very interesting new discovery from the {\it Suzaku}
observations is a marginal line feature close to 6.4~keV detected
in the XIS spectra of 3$\times$3 mode (Fig.~\ref{line}). This line
feature is only detectable in the FI detectors (XIS~0 and XIS~3)
and can not be seen in the spectrum of the BI detector (XIS~1);
this might be caused by the fact that the effective area of the BI
chip above 4~keV is smaller than that of the FI chips and thus the
XIS~1 is much less efficient in detecting line emission at higher
energy. Detailed analysis show that this line feature is at around
6.65~keV ($EW=150^{+300}_{-120}$~eV) and have a significance of
$\sim 4\sigma$. This line feature might be the K$_{\alpha}$
emission of highly ionised Fe. We do not find similar
emission feature in the spectra of the surrounding background.
This excludes the possibility that this line feature may be caused
by the background contamination, such as the intense diffuse
Galactic ridge emission. This line feature might
be originated from the surrounding ionised gas of the pulsar.

\section{Discussion}

We have presented the results of temporal and spectral analyses
for the X-ray emission from PSR~J1838--0655 observed by {\it
CHANDRA}, {\it RXTE} and {\it SUZAKU}. According to
Fig.~\ref{Profile}, the structure of the pulse profile of {\it
Suzaku}/PIN observation is very similar to that of {\it RXTE}/PCA.
There is some dissimilarity of the pulse profile derived by
\citet{AEDB2009} and ours. This may be caused by the different
selection of the zero epoch and the phase bins. We also divided
the photons of {\it RXTE} observations into 2--10~keV and
10--20~keV to plot the folded light curves (shown in
Fig.~\ref{RXTEProfile}), and we did not find obvious difference
from the profile structures.

Our spectral analysis indicates that there is a spectral break in
the X-ray spectra of the pulsar. This might be similar to the
spectral break discovered in the Crab pulsar \citep{Kup2001}. With
non-thermal emission model, the best fit photon indices are
$\Gamma_1\sim 1$ and $\Gamma_2\sim 1.5$ with a break energy
$E_c\sim 6.5$ keV and with the flux of $f\sim 1.6\times 10^{-11}$
ergs cm$^{-2}$ s$^{-1}$ in 2--20 keV. The results provide more
complete information of the spectral behaviour in addition to the
studies done by \cite{GH2008} and by \cite{AEDB2009}.
At a distance of $d\sim 6.6$ kpc, the efficiency of the energy conversion from the spin-down
luminosity, $L_p=5.5\times 10^{36}~\mathrm{ergs/s}$, to the X-ray
emissions is about 0.9\%, where we have used
$L_{0.8-10\mathrm{keV}}\sim 5.0\times 10^{34}~\mathrm{erg/s}$
estimated from our data analysis.

\begin{figure}
\centering
\includegraphics[width=9.0cm, height=8cm]{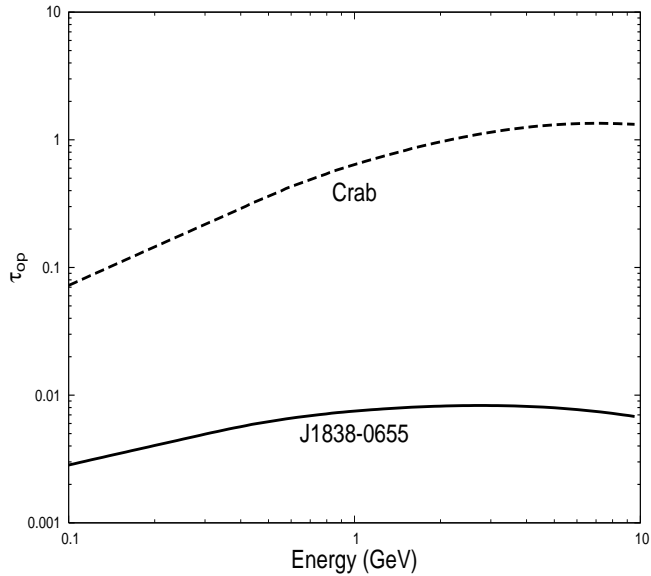}
\caption{\small Optical depths from the pair-creation of the
$\gamma$-ray photons as a function of energy. The solid line and
the dashed line represent the optical depths for PSR~J1838--0655
and the Crab pulsar, respectively.} \label{optics}
\end{figure}

If the observed X-ray emission is caused by a non-thermal process,
the emission might come from the magnetosphere around a particle
acceleration region, which is called a ``gap''. Comparing the
efficiency of the energy conversion from the spin-down energy to
the non-thermal X-ray emission with other known non-thermal X-ray
emitting pulsars listed in the tables 1. and 2. of \citet{KP2008}, we
find that the efficiency of PSR~J1838--0655, which was listed
in the table 3 of \citet{KP2008} for pulsars without known periods, is
the third largest and is larger than that of the Crab pulsar. One
might expect that the energy-conversion efficiency of PSR
J1838--0655 should be similar to that of the $\gamma$-ray pulsars,
PSRs~B0833--44 (the Vela) and B1706--44, because these three
pulsars have similar pulsar properties (e.g. the rotation periods
and the surface magnetic fields). However, the efficiencies of the
Vela pulsar and PSR~B1706--44 are about $5\times 10^{-4}$\% and
$3\times 10^{-3}$\%, respectively, and are much smaller than that
of PSR~J1838--0655. In fact, the efficiency of PSR~J1838--0655 is
in the range of younger $\gamma$-ray pulsars, such as the Crab
pulsar and PSR~B1509--58, which have
$L_{0.8-10\mathrm{keV}}/L_p\sim 0.34$\%, and $\sim 0.61$\%,
respectively. Therefore, PSR~J1838--0655 provides a unique
opportunity to investigate the non-thermal emitting process in the
pulsar magnetosphere and the connection between the X-ray and
$\gamma$-ray emission.

According to the outer magnetospheric emission model
\citep{Cheng2000}, the high efficiency of the Crab pulsar can be
explained as a result of the X-ray photons emitted via the
synchrotron radiation of the secondary electrons and positrons
produced in the pair-creation process of the primary
$\gamma$-rays, which are produced in the gap. In the gap of the
Crab pulsar, about 1\% of the spin-down energy is converted to the
$\gamma$-ray emission. Considering the optical depth of the
pair-creation of the $\gamma$-rays using the X-ray photon number
density outside the gap inferred from the Crab observations, we
find that most of the primary $\gamma$-ray photons above 1 GeV are
absorbed by the background X-ray photons before escaping from the
magnetosphere (Fig.~\ref{optics}). Because there are abundant new
born electron-positron pairs, which will emit X-ray photons via
the synchrotron radiation, the X-ray emission will dominate the
observed energy spectral distribution of the Crab pulsar and the
resultant conversion efficiency in the X-ray bands can be as large
as a few point percent.

The above emission model predicts the photon indices to be
$\Gamma_1\sim 1.5$ and $\Gamma_2\sim 2$ with a break at $E_c\sim
10$~keV \citep{TCC2007}. Roughly speaking, the predicted spectral
property is also similar to the observed properties of the X-ray
emission of PSR~J1838--0655, which shows a spectral break around
6.5~keV and has a high energy conversion efficiency. However, we
find that the X-ray emission model for the Crab pulsar can not be
applied to the case of PSR~J1838--0655 because the optical depth
of the pair-creation from the $\gamma$-rays is much smaller than
unity. Using the observed X-ray properties in this study, the
energy distribution of the X-ray photon number density in the
magnetosphere of PSR~J1838--0655 is
\begin{equation}
\frac{dN}{dE_X}=2.8\times 10^{22}d_{6.6\mathrm{kpc}}^2~\mathrm{cm^{-3} erg^{-1}}
\left\{
\begin{array}{@{\,}ll}
(E_x/6.5\mathrm{keV})^{-1}, \\
\mbox{$E_x\le 6.5\mathrm{keV}$}\\
(E_x/6.5\mathrm{keV})^{-1.5}, \\
\mbox{$E_x > 6.5\mathrm{keV}$}
\end{array}
\right.
\end{equation}
We have used $\frac{dN}{dE_X}=\frac{4\pi d^2}{c} \frac{dn}{dE_X}
{(\frac{R_{lc}}{2})}^{-2}$, where $\frac{dn}{dE_X}$ is the
observed number flux obtained in this paper, and $R_{lc}$ is the
light cylinder. Fig.~\ref{optics} compares the optical depths of
the pair-creation for the $\gamma$-ray photons of the
PSR~J1838--0655 (solid line) and of the Crab pulsar (dashed line)
before escaping the magnetosphere. We see that the optical depth
is much smaller than unity for the GeV $\gamma$-ray photons
emitted in the gap of PSR~J1838--0655, while it is about unity for
those of the Crab pulsar. So the absorption of the $\gamma$-ray
photons in the magnetosphere of PSR~J1838--0655 is very weak, and
only a few secondary pairs are created by the pair-creation
process with the background X-ray fields. As a result, the
synchrotron emission of the secondary pairs will not be able to
explain the flux of the observed X-ray emission. On this ground,
we conclude that the X-ray emission mechanism of the Crab pulsar
can not apply to PSR~J1838--0655, although the energy-conversion
efficiencies of both sources are similar.

A greatly inclined rotator, with the inclination angle $\alpha\sim
90^{\circ}$, is another possibility to explain the high energy
conversion in the X-ray emission of PSR~J1838--0655 and the
different efficiencies between PSR~J1838--0655 and the Vela
pulsar. With the outer magnetospheric model, the X-ray emission
from the Vela pulsar can be explained by the synchrotron emission
of the secondary pairs. The secondary pairs are produced by the
pair-creation process between the $\gamma$-rays and the background
X-ray field and/or by the magnetic pair-creation of the inwardly
propagating $\gamma$-rays, which pass through the region near the
stellar surface. The outer magnetospheric emission model studied
by \citet{ZJ2006} argued that the efficiency of the X-ray emission
increases with the inclination angle. This is caused by the fact
that (i) the strong acceleration of the outer gap model extends
between the null charge surface of the Goldreich-Julian charge
density and the light cylinder, and (ii) the null charge surface
shifts toward the stellar surface with increasing inclination
angles from $0^{\circ}$ to $90^{\circ}$. For the greater inclined
rotator, therefore, more $\gamma$-rays emitted inward directions
pass through vicinity of the stellar surface and are absorbed by
the strong magnetic field around the stellar surface. The created
secondary pairs can emit X-ray photons via the synchrotron
radiation.

The magnetic pair-creation condition for a photon with the energy
$E_{\gamma}$ may be written as
$(\frac{E_{\gamma}}{\mathrm{m_ec^2}})\frac{B}{B_c}\sin\theta_b\sim
0.2$ \citep{MH2003}, where $B_c=4.413\times 10^{13}$ Gauss and
$\theta_b$ is the angle between the magnetic field lines and the
propagating direction of $\gamma$-rays. Most of the pairs in the
gap are created around the null charge surface \citep{TSH2004}.
The distance from the stellar surface to the null charge surface
of the last-open field line is typically $r_n\sim
\frac{R_{lc}}{\sin^2(\theta_n-\alpha)}
{\sin^2(\theta_{lc}-\alpha)}\sin\theta_{lc}$, where
$\tan\theta_n=\frac{(3\tan\alpha+\sqrt{9\tan^2\alpha+8})}{2}$ and
$\tan\theta_{lc}=-\frac{(3+\sqrt{9+8\tan^2\alpha})}{4}\tan\alpha$
with $\theta_n$ being the angle of the null charge surface
measured from the north pole and $\theta_{lc}$ being the angle to
the point at which the last-open field line is tangent to the
light cylinder. If we use the radial distance $r_n$ as the
emission point, the closest distance from the stellar surface to
the trajectory of the inwardly propagating $\gamma$-rays emitted
on the null charge surface is $r_{min}\sim r_n\sin\theta_p$, where
$\theta_p$ is the angle of the null charge surface measured from
rotation axis. If we estimate the minimum inclination angle, above
which most of 1 GeV $\gamma$-rays are absorbed by the strong
magnetic field, with the magnetic pair-creation condition, we find
$\alpha>60^{\circ}$. In reality, the pairs are created above the
last-open field lines; for example, if we set the emission point
with $2r_n$, we would obtain $\alpha>70^{\circ}$.

\begin{figure}
\centering
\includegraphics[width=9.0cm, height=8cm]{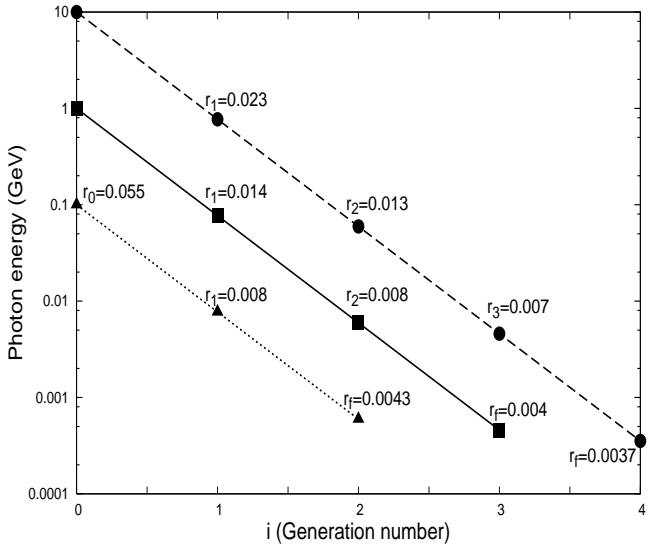}
\caption{\small Cascade of the magnetic pair-creation process. The vertical axis represents the energy of the emitted photons, and the
abscissa axis represents the generation number. The dashed, solid and
dotted lines represent the pair-cascade initiated by the photons with 10~GeV,
1~GeV and 0.1~GeV, respectively, emitted inwardly at the radial distance
$r_0=0.055$ in units of the light radius. The photon energies
of the generation number  $i$   represent the typical
energy of the synchrotron photons, and the values of
$r_i$ in the figure represent the
radial distances to the pair-creation positions of $i$-th generation.}
\label{cascade}
\end{figure}

The absorbed $\gamma$-ray photons are converted into
electron-positron pairs. The new born pairs lose their energy
rapidly via the synchrotron radiation; and the synchrotron photons
would be converted into electron-positron pairs via the magnetic
pair-creation process, if their energies are large enough for the
magnetic pair-creation process. The typical energy of the
synchrotron photons emitted by the first pairs are $E_1\sim0.075
E_{\gamma}$, where $E_{\gamma}$ is the energy of $\gamma$-ray
photons. The typical energy of the synchrotron photons emitted by
the $i$th generated pair can be written as
$E_i\sim(0.075)^{i}E_{\gamma}$.

If the magnetic fields on the path of the photons are strong
enough, the pair-creation cascade will continue until the typical
energy of the synchrotron photons becomes about 1~MeV. Let us take
the inclination angle of $80^{\circ}$ and a dipole magnetic field
as an example. The radial distance to the null charge surface on
the last-open field line in the magnetic meridional plane is
$r_n\sim 0.028R_{lc}$ (Figure~\ref{cascade}). We compute the
pair-creation cascade initiated by 1~GeV photons emitted inwardly
along the magnetic field line at the radial distance $r\sim
0.055R_{lc}$ and on the null charge surface in the magnetic
meridional plane. As summarized in Figure~\ref{cascade}, the first
pair-creation occurs at $r=0.014R_{lc}$ and the typical energy of
synchrotron photons emitted by the first generated pairs can be
written as $E_1\sim 75$~MeV. The sequence of the pair-creation
cascade stops after producing the 3rd generation pairs at $r_f\sim
0.004R_{lc}$. The 3rd generation pairs are produced with a Lorentz
factor of $\gamma\sim 6$ and emit synchrotron photons of $\sim
500$~keV. The spectrum of the emission from the pairs extends with
a photon index of $1.5-2$ (Cheng \& Zhang 1996) from $\sim
500$~keV to a break energy $E_{m}\sim 14~$keV, which is the
characteristic energy of the synchrotron photons from the
electrons/positrons with a Lorentz factor
$\gamma=1/\sin\theta_b\sim 1$. Below the break energy, the
spectrum will have a photon index of $\Gamma\sim 2/3$, which
represents the spectral slope of the synchrotron radiation below
its characteristic energy. We find this model can explain the
observed properties in the X-ray spectra of PSR~J1838--0655.

Therefore, if the X-rays from PSR~J1838--0655 are emitted in the
magnetosphere, an inclination angle close to $\alpha\sim
90^{\circ}$ would be preferred in order to explain the energy
conversion rate at X-ray bands. For the Vela pulsar, the energy
conversion rate can also be explained by $\alpha=30^{\circ}$ with
the magnetic pair-cascade model \citep{ZJ2006}. On this ground, it
is possible to explain the difference of the energy conversion
efficiencies of the Vela pulsar and PSR~B1838--0655 by the
difference of the inclination angles. The greatly inclined rotator model
may also predict the absence of the radio detection.
The copious pairs produced by the magnetic pair-creation near the pulsar can easily supply
the Goldreich-Julian number density. Therefore, if the magnetic pair-creation cascade initiated
by the $\gamma$-ray from the outer gap develops in the polar cap accelerator, the electric field
parallel to the magnetic field in the polar cap accelerator will be screened by the new born pairs,
indicating no coherent radio emission in the magnetosphere.

We note that most of $\gamma$-ray photons emitted outward can
escape from the magnetosphere without pair creations. The
outwardly migrating particles are accelerated using a
full potential drop along the field line before escaping the gap
from the outer boundary located around the light cylinder; on the
other hand, the inwardly migrating particles are accelerated with
a small part of the potential drop between the inner boundary and
the pair-creation position around the null charge surface. This implies
the luminosity of the outward $\gamma$-rays are one or
two-order larger than that of the inward $\gamma$-rays. If the
observed X-rays are originated from the inward
$\gamma$-rays, the outward $\gamma$-rays would be observed with a
flux similar to or greater than the X-ray flux, which is $f_x\sim
10^{-11}~\mathrm{ergs~cm^{-2}~s^{-1}}$. However, we note that the
observed $\gamma$-ray flux also depends on the viewing angle of
the observers. Therefore, future {\it Fermi} observations could
thus provide more information necessary to investigate the X-ray
emission process with a greatly inclined rotator model.


\section{Conclusions}

We have re-analysed the X-ray data of {\it CHANDRA}, {\it RXTE},
and {\it SUZAKU} to investigate the possible X-ray emission
mechanism for the X-ray pulsar PSR~J1838--0655. We find that a
broken power-law with the photon index varying from 1.0 to 1.5 and
with a break energy of $\sim$ 6.5 keV can provide a good fit for
the composite X-ray spectrum.
According to our results, we also found no significant energy
dependence of the pulse profiles.

We have discovered a marginal Fe K$_{\alpha}$ line feature in the
{\it Suzaku}/XIS data of PSR~J1838--0655. A similar line feature
was also detected in the radio pulsar
PSR~J1420--6048/AX~J1420.1--6049 \citep{RRJ2001} in {\it ASCA}
observations. Both pulsar are found to be associated with
high-energy TeV sources (i.e., HESS~J1420--607 for PSR~J1420--6048;
\citealp{Aha2006}); however, PSR~J1420--6048 not only has a radio
counterpart but also a hundred-MeV $\gamma$-ray one
(3EG~J1420--6048). These results might indicate that the Fe
K$_{\alpha}$ emission is associated with the TeV sources; however,
the relation between the line emission and the TeV sources is still not clear.

We have proposed and investigated the possible mechanism for the
observed X-ray emission of PSR~J1838--0655. The model for a
greatly inclined rotator is the most plausible scenario to explain
the high efficiency of the energy conversion from the spin-down
luminosity to X-ray emission. However, it is expected to accompany
with some level of $\gamma$-ray radiation. Because no unidentified
EGRET source exists in the region around PSR~J1838--0655, it
requires a deeper observation in the $\gamma$-ray bands (i.g.,
{\it Fermi} telescope) to solve the mystery of the X-ray emission
from this pulsar. This will also provide important clues to
understand how the X-rays and the $\gamma$-rays in the
magnetosphere of PSR~J1838--0655 are connected.

\section*{Acknowledgments}

The authors thank an anonymous referee for his/her helpful comments.
We also thank Dr. Albert Kong for the careful review of the manuscript
and thank Drs. Hsiang-Kuang Chang, Kwong-Sang Cheng and
Ronald Taam for fruitful discussion.
This research has made use of the data obtained through the High
Energy Astrophysics Science Archive Research Center Online
Service, provided by the NASA/Goddard Space Flight Center. This
work was partially supported by the National Science Council
through grants NSC 98-2811-M-008-044. CYH acknowledges support
from the National Science Council through grants
NSC~96-2112-M-008-017-MY3 and NSC~95-2923-M-008-001-MY3. JT was
supported by the Theoretical Institute for Advanced Reserch in
Astrophysics (TIARA), operated under Academia Sinica and the
National Science Council Excellence Projects program in Taiwan
through grant NSC 96-2752-M-007-007-PAE.

\label{lastpage}

\end{document}